\documentclass[10pt]{article}
\usepackage{epsf,amsmath,amssymb}
\usepackage[dvips]{graphicx}
\usepackage{citesort}

\DeclareMathOperator*{\doublesum}{\sum\sum}

\begin{document}
\title{
 Successive Phase Transitions in Antiferroelectric Liquid Crystal
 Systems \\
 -
 Axial Next-Nearest-Neighbor XY Model with Biquadratic Interaction
}
\date{}
\author{Masaya Koroishi, Masashi Torikai and Mamoru Yamashita\\
Department of Physics Engineering, 
Faculty of Engineering, Mie University\\
Kurimamachiya-cho 1577, Tsu, Mie, 514-8507, Japan
}

\maketitle

\begin{abstract}
  An axial next-nearest-neighbor XY model is studied as a model of
  chiral liquid crystals which exhibit many ferro-, ferri- and
  antiferroelectric tilted smectic phases.
  Depending on the values of interaction parameters, this model exhibits
  Ising symmetric (i.e., the tilt directions of directors are parallel
  or anti parallel) phases or XY symmetric phases.
  Phases with each type-of-symmetry show the character of devil's
  staircase, which has been observed in experiments.
\end{abstract}

\section{Introduction}
Several chiral liquid crystal (LC) compounds have found to exhibit
intermediate subphases, SmC${}^{*}_{\mathrm{FI1}}$ (also called
SmC${}^{*}_{\gamma}$), SmC${}^{*}_{\mathrm{FI2}}$ (also called AF), and
other ferrielectric phases (FI${}_\mathrm{H}$, FI${}_\mathrm{L}$ and FI)
between ferroelectric SmC${}^{*}$ and antiferroelectric
SmC${}^{*}_{\textrm{A}}$ phases~\cite{Isozaki1993,Matsumoto1994}. 
All these phases are tilted smectic phases and they have approximately
the same tilt angle from the layer normal.
Each phase consists of a helical stack of unit-set-of-layers (unit
cells);
the helix pitch is typically on the order of several
$\mu\mathrm{m}$~\cite{Hirst2002}.
Structures of SmC${}_{A}^{*}$, SmC${}^{*}_{\mathrm{FI1}}$ and
SmC${}^{*}_{\mathrm{FI2}}$ among those are
well-known~\cite{Matsumoto1994,Mach1998,Mach1999,Johnson2000,Cady2001}:
the unit cells characterizing SmC${}^{*}$, SmC${}_{A}^{*}$,
SmC${}^{*}_{\mathrm{FI1}}$, and SmC${}^{*}_{\mathrm{FI2}}$ consist of 1,
2, and 3, 4 layers,
respectively. 
The early experiments~\cite{Isozaki1993,Matsumoto1994,Fukuda1994} have
identified that the $\boldsymbol{c}$-directors (the projection of a director
onto layer-plane) of neighboring layers are parallel or antiparallel. 
This experimental result indicates that the system can be described with
an Ising model.
In fact, it has shown that an axial next nearest neighbor Ising (ANNNI)
model, which has competing interactions, exhibits infinitely many
subphases and can reproduce the transition sequence shown in
experiments~\cite{Yamashita1993,Yamashita1996a}.
However, the assumption of the Ising symmetry on these LC
systems is unnatural.
In practice, the $\boldsymbol{c}$-directors of layers are proved to deviate
slightly from the parallel-antiparallel configuration by later
high-resolution 
experiments~\cite{Mach1998,Mach1999,Johnson2000,Cady2001}.
In this context, various phenomenological theories have been carried out
to clarify the transition phenomena of the intermediate
phases~\cite{Lorman1994,Cepic1995,Wang1996,Roy1996,Skarabot1998}.
In these models a continuous rotational freedom of tilt direction is
taken into account.
To a greater or less extent, these models are generalization of ANNNI
model and the models are called axial next-nearest-neighbor XY (ANNNXY)
models.
Although the ANNNXY model exhibits infinitely many subphases, the system
does not undergo discrete transitions between these subphases but
continuously changes. 
By introducing an external aligning field on ANNNXY model, the
transitions between subphases are shown to be
recovered~\cite{Yamashita1999,Tanaka2000}.
However such external field is artificial and its origin is ambiguous.

As a more realistic model applicable to the antiferroelectric materials,
the ANNNXY model with a biquadratic interaction has been introduced
and introductory remarks on the crossover from XY character to Ising
character are given~\cite{Yamashita2000}.
In the present paper, we investigate the ANNNXY model with the
biquadratic interaction and disclose general properties of this model.

\section{ANNNXY Model with a Biquadratic Interaction}
Following ref.~\cite{Yamashita2000}, we introduce a Hamiltonian of the 
ANNNXY model with a biquadratic interaction as
\begin{equation}
 H = \sum_{l} H_{l},\\
\end{equation}
\begin{multline}
 H_{l} =
 - J \doublesum_{(i, j)} \cos\left(\varphi^{(i)}_{l} - \varphi^{(j)}_{l}\right)
 - J_{1} \sum_{i} \cos\left(\varphi^{(i)}_{l+1} - \varphi^{(i)}_{l}\right)\\
 - J_{2} \sum_{i} \cos\left(\varphi^{(i)}_{l+2} - \varphi^{(i)}_{l}\right)
 - K \sum_{i} \cos\left[2\left(\varphi^{(i)}_{l+1} - \varphi^{(i)}_{l}\right)\right],
 \label{eq:Hamiltonian}
\end{multline}
where the $\varphi^{(i)}_{l}$ denotes the azimuthal angle of a molecule
$i$ in $l$-th layer.
The parameter $J (>0)$ is the interaction parameter between molecules
belonging to the same layer.
The summation in the first term of eq.\eqref{eq:Hamiltonian} is done
over all neighboring pairs in the $l$-th layer.
The parameters $J_{1}$ and $J_{2}$ are the interaction parameters
between molecules in the nearest neighboring layers and next nearest
neighboring layers, respectively.
Let $J_{1}$ be positive without loss of generality since the
transformation $J_{1} \to -J_{1}$ followed by
$\{\varphi^{(i)}_{l}\} \to \{\varphi^{(i)}_{l} + \pi l \}$ leaves the
above Hamiltonian invariant.
In the following, we set $J_{2}$ to be negative.
The last term of eq.\eqref{eq:Hamiltonian} is the biquadratic
interaction with an interaction parameter $K (>0)$.
This biquadratic interaction term stabilizes the parallel and
antiparallel configuration of $\boldsymbol{c}$-directors.
This Hamiltonian can be viewed as a truncated Fourier approximation of
an exact Hamiltonian.

Our model turns to the ordinary ANNNXY model for vanishing value of $K$.
In the limit of $K \to \infty$, it turns to the ANNNI model.
Thus for sufficiently large $K$ the ordered phase has parallel and
antiparallel $\boldsymbol{c}$-directors;
we call such phases Ising symmetric phases.
We call the other phases XY symmetric phases. 

In the present paper, we ignore the helical structure of the chiral
smectic phases, and assume these phases as simple periodic repetitions
of unit cells. 
Thus a smectic phase is characterized by a set of order parameters in
the unit cell, and may be labeled with its wavenumber $q$.
We note that the transformation
$\{\varphi_{l} \to \varphi_{l} + \pi l\}$ results in $q \to 1/2-q$.
Thus, e.g., $q=1/3$ (SmC${}^{*}_{\mathrm{FI1}}$) and $q=1/2$
(SmC${}^{*}_{A}$) phases under negative $J_{1}$ correspond,
respectively, to $q=1/6$ and $q=0$ phases under positive $J_{1}$. 

At zero temperature, there are three stable phases~\cite{Yamashita2000}:
$q=0$ phase, $q=1/4$ phase, and a phase Q in which the wave number
changes continuously in the range $0 < q < 1/4$.
The phase diagram for temperature $T=0$ is shown in the inset of
Fig.~\ref{fig:K0phaseDiagram}. 
In the phase $q=1/4$, irrespective of the value $K$, the azimuthal 
angles in a unit cell can be written as
$\varphi_{1}=\varphi_{2}=0$ and $\varphi_{3}=\varphi_{4}=\pi$, i.e., 
this phase has Ising symmetry.
The stable distribution of azimuthal angle in phase Q is
$\varphi_{l}=2\pi ql$, and the stable wave number is determined by
interaction parameters as $\cos 2\pi q = J_{1}/4(J_{2}+K)$.

\section{Theoretical Methods and Results}
\subsection{Mean Field Approximation}
In this section, we investigate the stability of phases at finite
temperature within the mean field approximation.
We choose the order parameters of $l$-th layer as
\begin{equation}
 \begin{split}
 c_{l} &= \left\langle \cos \varphi^{(i)}_{l} \right\rangle , \\
 s_{l} &= \left\langle \sin \varphi^{(i)}_{l} \right\rangle , \\
 C_{l} &= \left\langle \cos 2\varphi^{(i)}_{l} \right\rangle, \\
 S_{l} &= \left\langle \sin 2\varphi^{(i)}_{l} \right\rangle,
 \end{split} \label{eq:orderParameters}
\end{equation}
where the angular brackets indicate a thermal average over all molecules
in a single layer.
We can assume, without loss of generality, that in Ising symmetric
phases the $\boldsymbol{c}$-directors are parallel to $x$-axis and thus
$s_{l}=S_{l}=0$. 
The mean fields due to these order parameters are
\begin{equation}
 \begin{split}
  \xi_{l}  &= \beta \left[Jzc_{l} + J_{1}(c_{l-1} + c_{l+1})
                         + J_{2}(c_{l-2} + c_{l+2})\right], \\
  \eta_{l} &= \beta \left[Jzs_{l} + J_{1}(s_{l-1} + s_{l+1})
                         + J_{2}(s_{l-2} + s_{l+2})\right],\\
  \mu_{l}  &= \beta K(C_{l-1} + C_{l+1}),\\
  \nu_{l}  &= \beta K(S_{l-1} + S_{l+1}),
 \end{split} \label{eq:meanField}
\end{equation}
where we assumed that each molecule interacts $z$ molecules in the same
layer, one molecule in each of the two nearest neighboring layers, and
one molecule in each of the two next nearest neighboring layers.
We can derive the mean field free energy per layer~\cite{Yamashita2000}
using the above order parameters and mean fields as
\begin{multline}
  \Phi
  = \frac{1}{p} \sum_{l}^{p}
  \biggl[
  \frac{1}{2}
    \left\{
     Jz c_{l} + J_{1}(c_{l-1} + c_{l+1}) + J_{2}(c_{l-2} + c_{l+2})
    \right\}c_{l}\\
    + \frac{1}{2}
    \left\{
     Jz s_{l} + J_{1}(s_{l-1} + s_{l+1}) + J_{2}(s_{l-2} + s_{l+2})
    \right\}s_{l} \\
    + \frac{1}{2} K (C_{l-1} + C_{l+1})C_{l}
    + \frac{1}{2} K (S_{l-1} + S_{l+1})S_{l}  \\
    - \frac{1}{\beta}
  \ln \tau(\xi_{l}, \eta_{l}, \mu_{l}, \nu_{l}) \biggr],
\label{eq:freeEnergy}
\end{multline}
where $p$ is a period (number of layers in a unit cell).
The function $\tau$ in the last term of eq.\eqref{eq:freeEnergy} is the
one-molecule partition function defined as 
\begin{equation}
 \tau(\xi_{l}, \eta_{l}, \mu_{l}, \nu_{l}) =
  \int_{0}^{2\pi} 
  \exp \left(
               \xi_{l} \cos\varphi
             + \eta_{l} \sin\varphi
             + \mu_{l} \cos 2\varphi
             + \nu_{l} \sin 2\varphi
  \right) d\varphi.
\end{equation}
From the derivatives of the free energy with respect to the mean fields,
we obtain a set of self-consistent equations:
\begin{equation}
 \begin{split}
  c_{l} & =
        \frac{\partial}{\partial \xi_{l}}
  \ln \tau(\xi_{l}, \eta_{l}, \mu_{l}, \nu_{l}), \\
  s_{l} & =
        \frac{\partial}{\partial \eta_{l}}
  \ln \tau(\xi_{l}, \eta_{l}, \mu_{l}, \nu_{l}), \\
  C_{l} & =
        \frac{\partial}{\partial \mu_{l}}
  \ln \tau(\xi_{l}, \eta_{l}, \mu_{l}, \nu_{l}), \\
  S_{l} & =
        \frac{\partial}{\partial \nu_{l}}
  \ln \tau(\xi_{l}, \eta_{l}, \mu_{l}, \nu_{l}),
 \end{split} \label{eq:self_consistent2}
\end{equation}
where $\xi_{l}$, $\eta_{l}$, $\mu_{l}$, and $\nu_{l}$ in the right-hand
sides are functions of order parameters defined in
eqs.\eqref{eq:meanField}.
One of the solutions of eqs.\eqref{eq:self_consistent2} which minimizes
the free energy gives a set of order parameters of the thermodynamically
stable state.

The transition temperature between ordered and disordered phases is
\begin{equation}
 T_{\mathrm{c}}
  =
  \left(1 + P_{1}(\zeta_{\mathrm{c}}) \right) T_{\mathrm{c}}^{\mathrm{XY}},
  \label{eq:Tc}
\end{equation}
where $T_{\mathrm{c}}^{\mathrm{XY}}$ and $P_{1}(z)$ are, respectively, 
\begin{gather}
 T_{\mathrm{c}}^{\mathrm{XY}}
  =
  \frac{1}{2}Jz + \frac{J_{1}{}^{2}}{8|J_{2}|} + |J_{2}|,\\
 P_{1}(z)
  =
  I_{1}(z)/I_{0}(z), \label{eq:TcXY}
\end{gather}
with the $n$-th order modified Bessel functions $I_{n}(\zeta)$.
The variable $\zeta_{\mathrm{c}}$ is a solution of
\begin{equation}
 \zeta = 2 K P_{1}(\zeta)/T_{\mathrm{c}}, \label{eq:zetaC}
\end{equation}
thus $T_{\mathrm{c}}$ is given by solving
eqs.\eqref{eq:Tc} and \eqref{eq:zetaC} self-consistently.
The above critical temperature is derived~\cite{Yamashita2000}, under
the assumption that the transition is second-order, by expanding the
free energy $\Phi(c_{l}, s_{l}, C_{l}, S_{l})$ around $c_{l}=s_{l}=0$,
$C_{l}=C^{(0)}$ and $S_{l}=S^{(0)}$, and letting the quadratic terms of
the expansion vanish at critical temperature.
The order-disorder transition temperature deviates from eq.\eqref{eq:Tc}
if the transition is first-order.
Indeed, as we will see below, the transition temperature is slightly
higher than $T_{\mathrm{c}}$ when $K$ is comparable to
$T_{\mathrm{c}}$.

In the following, we restrict our consideration to a set of
\textit{major} ordered phases with wavenumbers $q=0$, $1/8$, $1/6$,
$1/5$, $1/4$, and a disordered phase.
Since we discuss the relative stability of these phases rather than
infinitely many phases, the resulting phase diagrams should be viewed as
rough approximations of correct phase diagrams. 
However, we can expect, from the analogy of ANNNI model, that other than
the above six phases will occupy relatively small regions on the phase
diagrams, and that we can safely ignore these minor 
phases.

In order to reduce the number of order parameters, we make a reasonable
assumption that the ordered phases are highly symmetric.
In precise, when the period $p$ is odd (i.e., $p=5$ in this paper), we
assume that a set of $\boldsymbol{c}$-directors of a unit cell has
mirror symmetry with respect to a plane perpendicular to the smectic
layers; 
when $p$ is even, we assume a two-fold rotational symmetry of the set of
$\boldsymbol{c}$-directors, in addition to the mirror symmetry.

In the following calculation, we set $J=J_{1}$ and $z=6$.
We take $|J_{2}|$ as the unit of energy.
Each ordered phase has a set of (reduced number of) self-consistent
equations \eqref{eq:self_consistent2}.
We solved the sets of equations numerically, and determined the
thermodynamic stable phase by comparing the free energy per molecule.

\subsection{Phase Diagrams}
\begin{figure}
 \begin{center}
  \includegraphics[width=8.cm,keepaspectratio,clip]{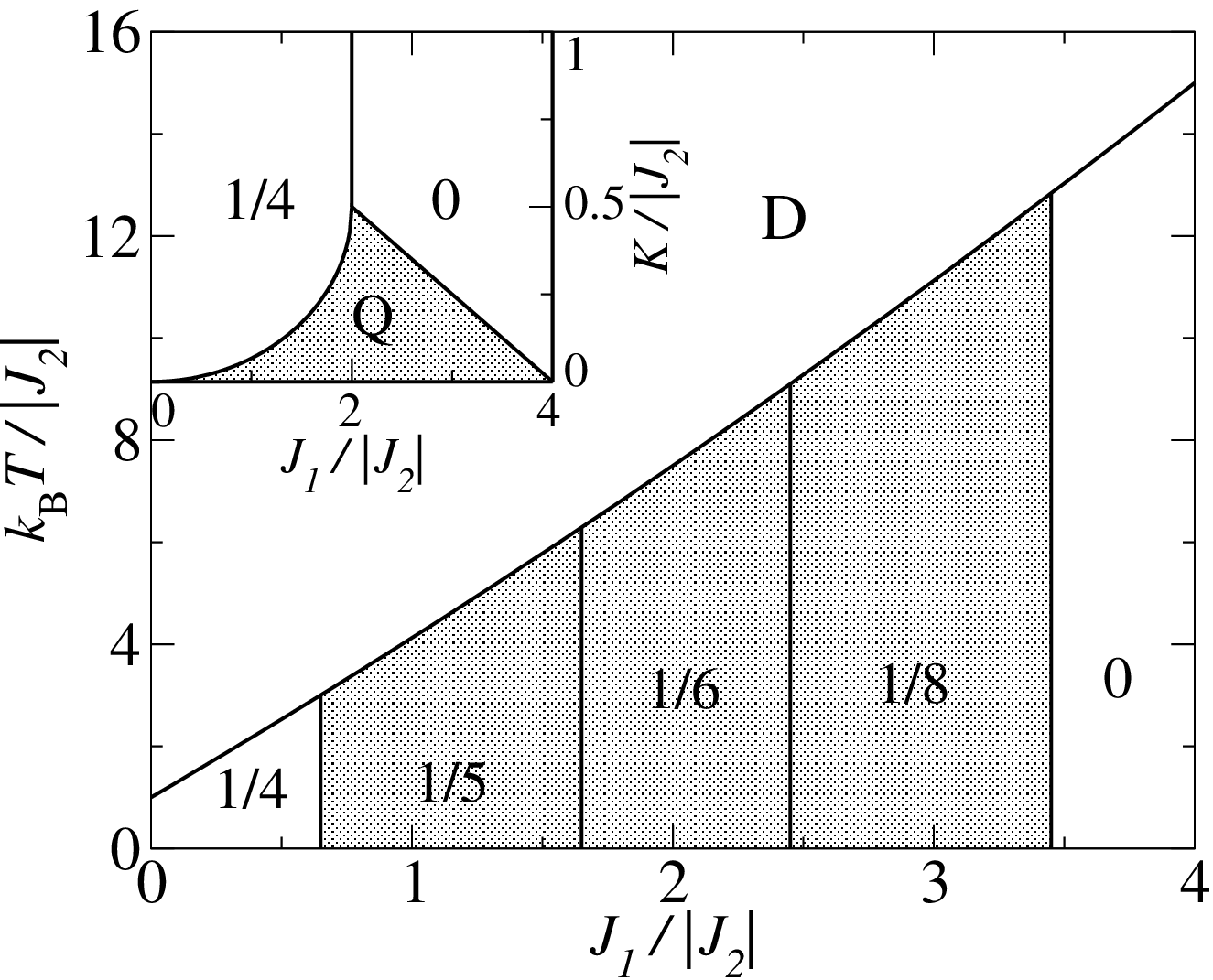}
 \end{center}
 \caption{
 Phase diagram in the $T$-$J_{1}$ plane for $K=0$.
 Each fractional number indicates the wavenumber $q$ of relatively
 stable phase, and D denotes a disordered phase.
 The shaded areas indicate the phases with XY symmetry, where
 non-vanishing $s_{l}$ and $S_{l}$ are allowed;
 while the unshaded areas with $q=0$ and $1/4$ indicate the phases with
 Ising symmetry, where $s_{l}=S_{l}=0$.
 The inset is the phase diagram in the $K$-$J_{1}$ plane for zero
 temperature.
 The wavenumber of the phase Q continuously changes from 0 to 1/4.
 }
 \label{fig:K0phaseDiagram}
\end{figure}

At $K=0$, our model is reduced to an ordinary ANNNXY model.
Then the system transits through the phases of $0 \le q \le 1/4$
continuously~\cite{Lorman1994,Cepic1995,Wang1996,Roy1996,Skarabot1998}
and thus the successive phase transitions are never observed.
We show the phase diagram for $K=0$ in Fig.~\ref{fig:K0phaseDiagram},
in which we considered the relative stability of $q=0$, $1/8$, $1/6$,
$1/5$, and $1/4$ phases.
The phases with $q=1/8$, $1/6$, and $1/5$ have XY symmetry, while $q=0$
and $1/4$ phases have Ising symmetry.
It seems there is a contradiction between $K$-$J_{1}$ diagram (inset)
and $T$-$J_{1}$ diagram at $K=0$ and $T=0$, i.e., stable regions of
$q=0$ and $1/4$ phases are overestimated in $T$-$J_{1}$ diagram;
such a contradiction, which can be observed in all the following phase
diagrams, is due to our approximation in which infinitely many minor
phases are ignored.

\begin{figure}
 \begin{center}
  \includegraphics[width=8.cm,keepaspectratio,clip]{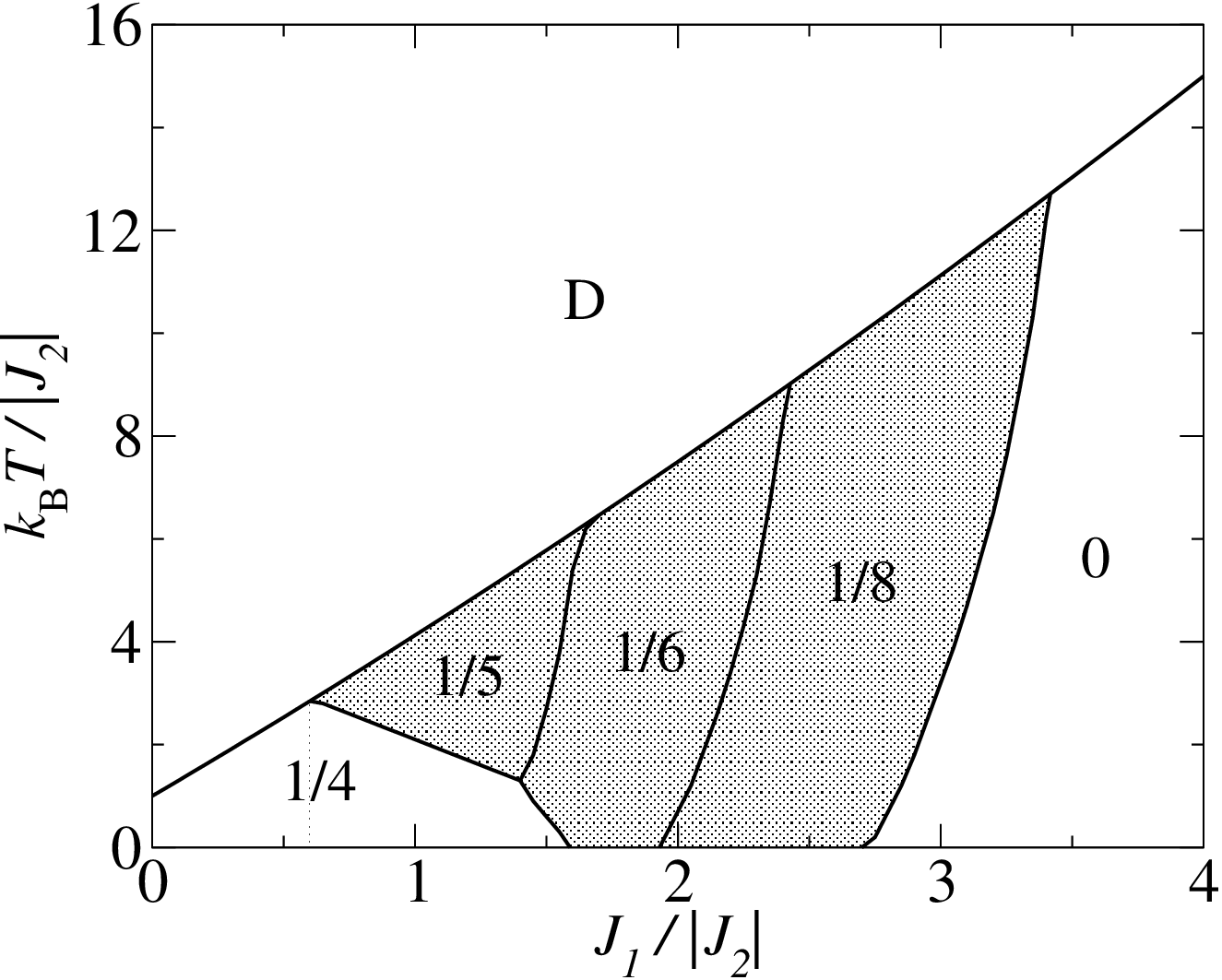}
 \end{center}
 \caption{
 The same as in Fig.~\ref{fig:K0phaseDiagram} but for $K=0.2$.
 }
 \label{fig:K02phaseDiagram}
\end{figure}

Since the effect of the biquadratic interaction stabilizes the Ising
symmetric phases, $q=0$ and $1/4$ phases dominate larger regions as $K$
increases.
The phase diagram for $K=0.2$, Fig.~\ref{fig:K02phaseDiagram}, shows
that the Ising symmetric phases spread from the lower-temperature region.
The phases with $q=1/8$, $q=1/6$, and $q=1/5$ still have XY symmetry.

\begin{figure}
 \begin{center}
  \includegraphics[width=8.cm,keepaspectratio,clip]{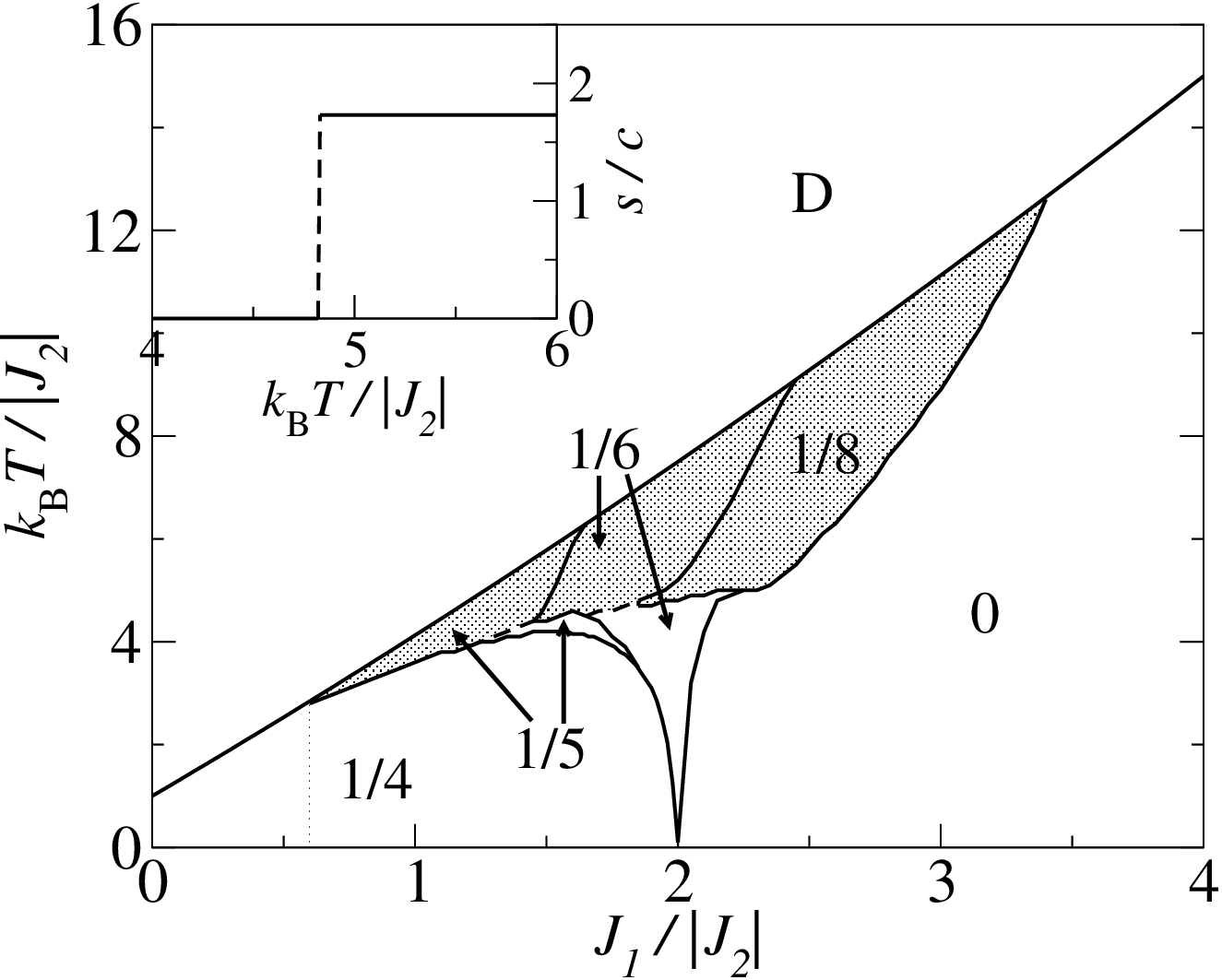}
 \end{center}
 \caption{
 The same as in Fig.~\ref{fig:K0phaseDiagram} but for $K=1.0$.
 The phases with $q=1/6$ and $1/5$ exhibit Ising symmetric phases.
 The broken line indicates the boundary between XY symmetric phases and
 Ising symmetric phases.
 (Inset): The order parameter ratio $s/c$ of $q=1/6$ phase as a function
 of temperature at $J_{1}/|J_{2}|=1.8$ around the XY-Ising transition
 temperature.
 }
 \label{fig:K10phaseDiagram}
\end{figure}

When $K/|J_{2}|$ exceeds $0.5$, the phases with $0 < q < 1/4$ vanish for
any $J_{1}$ at zero-temperature (see the inset of
Fig.~\ref{fig:K0phaseDiagram}). 
The phases with $q=1/8$, $1/6$, and $1/5$ change qualitatively at this
stage. 
In these phases, Ising symmetric phases appear at lower temperature
region and the Ising symmetric regions expand as $K$ increases.
Figure~\ref{fig:K10phaseDiagram} is a phase diagram for $K=1.0$.
In this phase diagram, we can clearly observe the cross-over from XY
symmetric phases to Ising symmetric phases.

\begin{figure}
 \begin{center}
  \includegraphics[width=8.cm,keepaspectratio,clip]{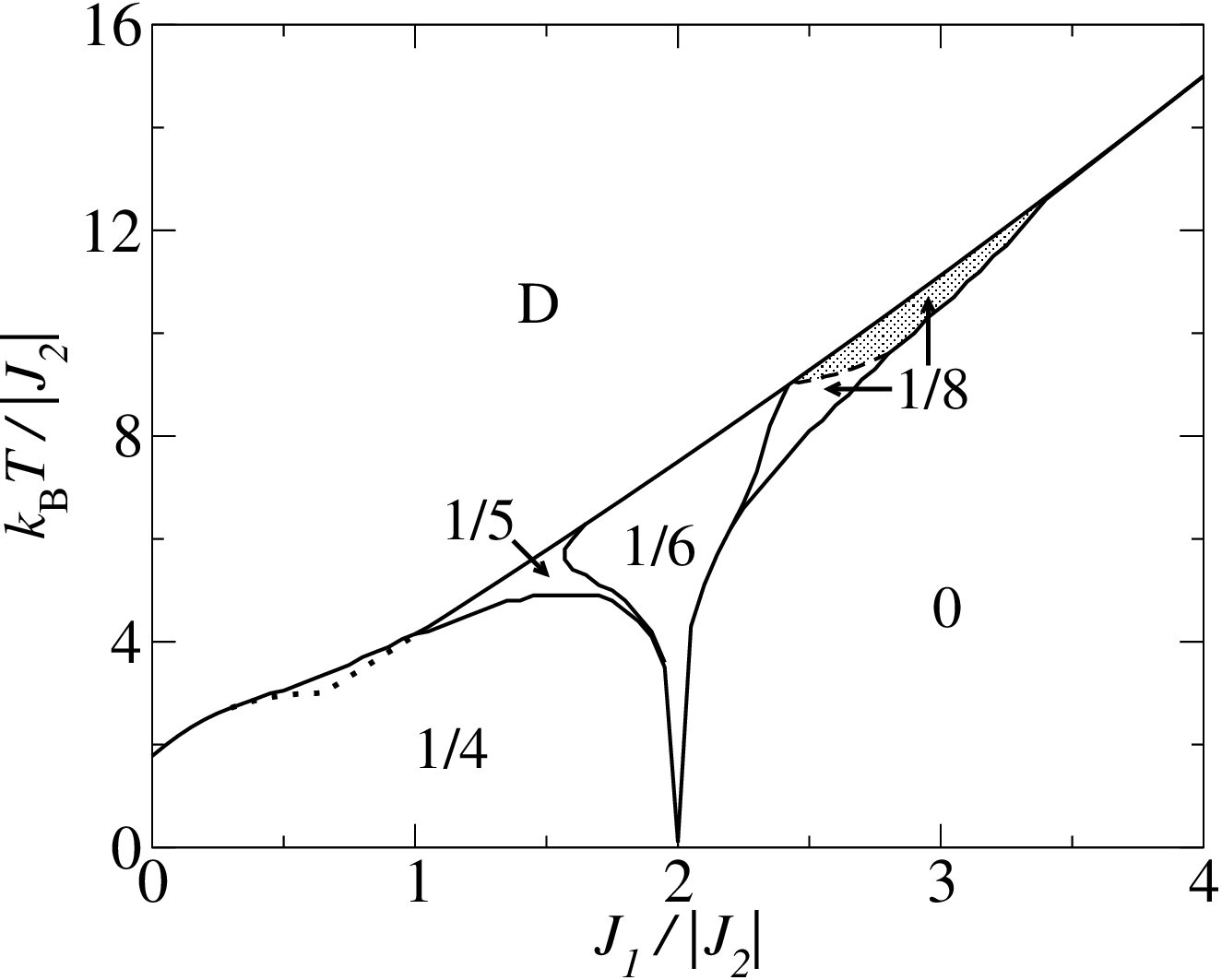}
 \end{center}
 \caption{
 The same as in Fig.~\ref{fig:K10phaseDiagram} but for $K=3.0$.
 The XY symmetric phase remains only in $q=1/8$ phase.
 The dotted line around $J_{1}/|J_{2}| \simeq 0.5$ and
 $k_{\mathrm{B}}T/|J_{2}| \simeq K = 3$ is $T_{\mathrm{c}}$ defined in
 eq.\eqref{eq:Tc}.
 The order($q=1/4$)-disorder transition temperature (indicated by solid
 line) is slightly higher than the $T_{\mathrm{c}}$ and the transition
 is first-order in the region where $K$ and
 $k_{\mathrm{B}}T_{\mathrm{c}}/|J_{2}|$ are comparable.
 }
 \label{fig:K30phaseDiagram}
\end{figure}

Figure~\ref{fig:K30phaseDiagram} shows the phase diagram for $K=3.0$.
The Ising symmetric phases are spread out almost the whole region.
This figure also shows a difference from Figs.~\ref{fig:K0phaseDiagram},
\ref{fig:K02phaseDiagram} and \ref{fig:K10phaseDiagram} in the
order-disorder transition temperature. 
The temperature $T_{\mathrm{c}}$ (the order-disorder transition
temperature under the assumption of the second-order transition)
deviates from $T_{\mathrm{c}}^{\mathrm{XY}}$ when $K$ exceeds unity (see
eqs.\eqref{eq:Tc} and \eqref{eq:zetaC}); 
the deviation arises from the region where
$k_{\mathrm{B}}T_{\mathrm{c}}/|J_{2}|<K$.
However the $T_{\mathrm{c}}$ does not give correct order-disorder
transition temperature.
In fact, the actual transition temperature is slightly higher than
the calculated $T_{\mathrm{c}}$ in the region where $K$ and
$k_{\mathrm{B}}T_{\mathrm{c}}/|J_{2}|$ are comparable, as shown in
Fig.~\ref{fig:K30phaseDiagram}. 
This fact indicates that the order-disorder transition is first-order,
rather than second-order.
Indeed, we observed that the order parameters changes discontinuously at
the region where the order-disorder transition temperature is different
from $T_{\mathrm{c}}$.

\section{Summary and Discussions}
We studied the transition behavior of ANNNXY model with biquadratic
interaction which can be applicable to the successive phase transitions
of antiferroelectric smectics;
the biquadratic interaction is interpreted as the second Fourier
component of a pair directional interaction.
The crossover from the XY symmetry to the Ising one is identified
for several representative phases.
The Ising symmetric phase has been suggested to appear from the low
temperature region as the strength of biquadratic interaction is
increased~\cite{Yamashita2000}. 
At the XY symmetric phase with period $p$, the difference of tilt
directions of directors in successive layers,
$\varphi_{l+1}-\varphi_{l}$, is shown to be $2\pi/p$, which indicates a
simple helical structure. 

Our results show that the ANNNXY model with biquadratic interaction
reproduce the phase sequence observed in the experiments.
However, we may not expect the quantitative agreement with experiments.
As an example, let us consider the LC molecules in $q=1/6$
phase, which is essentially equivalent to $q=1/3$ phase for negative
$J_{1}$, and thus this phase corresponds to the 3-layer
SmC${}^{*}_{\mathrm{FI1}}$. 
Ellipsometric experiments have shown that $\boldsymbol{c}$-directors of
this phase do approximately but not exactly lie in a
co-plain~\cite{Johnson2000}.
In precise, the azimuthal angles of molecules in a period are
$-\delta$, $\delta$, and $\pi$, with $\delta \lesssim 0.5$rad
($\sim 30^{\circ}$).
In terms of our model, this fact corresponds to the situation where
$|s_{l} / c_{l}| \lesssim 0.5$.
The inset of Fig.~\ref{fig:K10phaseDiagram} shows a typical behavior
of $ s_{l} / c_{l} $ in $q=1/6$ phase at the XY-Ising transition
temperature. 
As shown in Fig.~\ref{fig:K10phaseDiagram}, the ratio $ s_{l} / c_{l} $
is rather large($\sim 1.7$) at high temperature and approximately zero
at low temperature;
we cannot find any region where $s_{l} / c_{l}$ is comparable to the
experimental results.
For quantitative agreement of the present results with experimental
ones, some additional modification to the present model is required.
We may expect to obtain quantitatively reliable results by introducing a
chiral interaction together with long range
interactions~\cite{Yamashita1993,Yamashita1996b,Cepic2001}, as done by
Olson \textit{et al.}~\cite{Olson2002} to explain the $\delta$ in the
framework of phenomenological free energy. 

The profile of phase diagram for Ising symmetric phases in
Fig.~\ref{fig:K30phaseDiagram} suggests apparently a character of
devil's staircase~\cite{Yamashita1993,Yamashita1996a} and thus the
system undergoes successive phase transitions.
For phases of XY symmetry in Figs.~\ref{fig:K02phaseDiagram} and
\ref{fig:K10phaseDiagram}, however, it is not clear from our present
analysis whether the system exhibits the successive phase transitions or
not.
Analysis of the stability of the soliton excitation can be a powerful
tool to certify the successive transitions between XY symmetric
phases.
In the ANNNXY model under the two-fold external field, the character of
devil's staircase is certified even in the ground
state~\cite{Yamashita1999}.
However, in the present model, the phase changes continuously at the
ground state for small biquadratic interaction as shown in the inset of
Fig.~\ref{fig:K0phaseDiagram}. 
In this respect, the phase boundary of main phase should be studied at
finite temperature, which will disclose the character of successive
phase transitions.


\begin{thebibliography}{99}
 \bibitem{Isozaki1993}
         T. Isozaki, T. Fujikawa, H. Takezoe, A. Fukuda, T. Hagiwara,
         Y. Suzuki and I. Kawamura,
         Phys. Rev. B \textbf{48} (1993) 13439.
 \bibitem{Matsumoto1994}
         T. Matsumoto, A. Fukuda, M. Johno, Y. Motoyama, T. Yui,
         S. S. Seomun and M. Yamashita,
         J. Mater. Chem. \textbf{9} (1999) 2051.
 \bibitem{Hirst2002}
         L. S. Hirst, S. J. Watson, H. F. Gleeson,
         P. Cluzeau, P. Barois, R. Pindak, J. Pitney, A. Cady,
         P. M. Johnson, C. C. Huang, A-M. Levelut, G. Srajer,
         J. Pollmann, W. Caliebe, A. Seed, M. R. Herbert, J. W. Goodby,
         and M. Hird, 
         Phys. Rev. E \textbf{65} (2002) 041705.
 \bibitem{Mach1998}
         P. Mach, R. Pindak, A.-M. Levelut, P. Barois,
         H. T. Nguyen, C. C. Huang, and L. Furenlid,
         Phys. Rev. Lett. \textbf{81} (1998) 1015.
 \bibitem{Mach1999}
         P. Mach, R. Pindak, A.-M. Levelut, P. Barois,
         H. T. Nguyen, H. Baltes, M. Hird, K. Toyne, A. Seed,
         J. W. Goodby, C. C. Huang, and L. Furenlid,
         Phys. Rev. E \textbf{60} (1999) 6793. 
 \bibitem{Johnson2000}
         P. M. Johnson, D. A. Olson, S. Pankratz,
         T. Nguyen, J. Goodby, M. Hird, and C. C. Huang,
         Phys. Rev. Lett. \textbf{84} (2000) 4870.
 \bibitem{Cady2001}
         A. Cady, J. A. Pitney, R. Pindak, L. S. Matkin, S. J. Watson,
         H. F. Gleeson, P. Cluzeau, P. Barois, A.-M. Levelut,
         W. Caliebe, J. W. Goodby, M. Hird, and C. C. Huang,
         Phys. Rev. E \textbf{64} (2001) 050702.
 \bibitem{Fukuda1994}
         A. Fukuda, Y. Takanishi, T. Isozaki, K. Ishikawa and
         H. Takezoe,
         J. Mater. Chem. \textbf{4} (1994) 997.
 \bibitem{Yamashita1993}
         M. Yamashita and S. Miyazima,
         Ferroelectrics \textbf{148} (1993) 1.
 \bibitem{Yamashita1996a}
         M. Yamashita,
         Ferroelectrics \textbf{181} (1996) 201.
 \bibitem{Yamashita1999}
         M. Yamashita and S. Takeno,
         J. Phys. Soc. Jpn. \textbf{68} (1999) 1493.
 \bibitem{Lorman1994}
         V. L. Lorman, A. A. Bulbitch and P. Toredano,
         Phys. Rev. E \textbf{49} (1994) 1369.
 \bibitem{Cepic1995}
         M. Cepic and B. Zeks,
         Mol. Cryst. Liq. Cryst. \textbf{263} (1995) 61.
 \bibitem{Wang1996}
         X. Y. Wang and P. L. Taylor,
         Phys. Rev. Lett. \textbf{76} (1996) 640.
 \bibitem{Roy1996}
         A. Roy and N. V. Madhusudana,
         Europhys. Lett. \textbf{36} (1996) 22.
 \bibitem{Skarabot1998}
         M. Skarabot, Mojca Cepic, B. Zeks, R. Blinc, G. Heppke,
         A. V. Kityk, and I. Musevic,
         Phys. Rev. E \textbf{58} (1998) 575.
 \bibitem{Tanaka2000}
         S. Tanaka and M. Yamashita,
         Ferroelectrics \textbf{245} (2000) 209.
 \bibitem{Yamashita2000}
         M. Yamashita and S. Tanaka,
         Ferroelectrics \textbf{245} (2000) 217.
 \bibitem{Yamashita1996b}
         M. Yamashita,
         J. Phys. Soc. Jpn. \textbf{65} (1996) 2122.
 \bibitem{Cepic2001}
         M. Cepic and B. Zeks,
         Phys. Rev. Lett. \textbf{87} (2001) 85501.
 \bibitem{Olson2002}
         D. A. Olson, X. F. Han, A. Cady, and C. C. Huang,
         Phys. Rev. E \textbf{66} (2002) 021702.
\end{thebibliography}
\end{document}